\documentstyle[12pt]{article}

\def \be{\begin{equation}}
\def \ee{\end{equation}}
\def \bea{\begin{eqnarray}}
\def \eea{\end{eqnarray}}
\def \f{\frac}
\def \n{\noindent}

\def \ra{\rightarrow}
\def \g{\gamma}
\def \cdg{c^{\g}_d}
\def \cdz{c^Z_d}
\def \cvg{c^{\g}_v}
\def \cag{c^{\g}_a}

\def \caz{c^Z_a}

\begin{document}

\begin{flushright}
PRL-TH-94/31
\end{flushright}

\def\baselinestretch{2}
\vskip 5mm

\begin{center}
{\large \bf CP-violating asymmetries in $e^+e^-\ra t\bar{t}$ 
 with longitudinally polarized electrons}\\[3mm]
{\large P. Poulose and Saurabh D. Rindani}\\[2mm]
{\it Theory Group, Physical Research Laboratory}\\
{\it Navrangpura, Ahmedabad  380 009, India}
\end{center}

\begin{abstract}
New CP-violating asymmetries of decay leptons in
$e^+\,e^-\;\ra\;t\,\bar{t}$, arising from electric and weak dipole
couplings of $t\,\bar{t}$ to $\gamma$ and $Z$, are examined in the
case of unpolarized and longitudinally polarized electrons. The
new asymmetries measured together with the old ones can help to
determine independently the real and imaginary parts of the
electric as well as  weak dipole couplings.
Longitudinal beam polarization, if present, obviates the need
for the simultaneous measurement of more than one asymmetry, and
enhances considerably the sensitivity to the CP-violating
parameters.  Numerical results are presented for the Next Linear
Collider with $\sqrt{s}=500$ GeV and
$\int{\cal L}\,dt\;=\;10\,{\rm fb}^{-1}$. 

\end{abstract}

\newpage

Experimental results at the $p\bar{p}$ collider Tevatron at
Fermilab [1] indicate that the top quark is heavy
($m_t=174\pm 10^{+13}_{-12}\;{\rm GeV})$, as anticipated from the LEP results
and one-loop radiative corrections in the standard model (SM).
It has been suggested that a heavy top ($m_t >$ 120 GeV) would
decay before it can hadronize [2], and therefore its decay
products would preserve useful information on its polarization.
This information could be utilized, for example, in
investigating possible CP violation in $t\,\bar{t}$ production in
hadronic and $e^+\,e^-$ collisions [3-7]. 

 At the present time,
CP violation has only been seen in the K-meson system. While
these observations are consistent with CP violation arising from a single
phase in the Cabibbo-Kobayashi-Maskawa matrix in SM, there are
several extensions of SM which can accommodate the observed CP violation.
Some of these models predict enhanced CP violation for a heavy
top quark [4,5], which could show up in a CP-violating electric-dipole type
$t\,\bar{t}\,\gamma$ vertex and an analogous ``weak"-dipole
$t\,\bar{t}\,Z$ vertex.  An observation of these would be
interesting and useful from the theoretical point of view.

Proposals have been recently forwarded on using top
polarization asymmetry, and subsequent asymmetry in the decay
distributions [3,5,6], to measure  CP violation in top
production\footnote{ CP violation in top decays has also been
considered. See, for example, Ref. [8].}.  The
measurement of CP-odd correlations among the momenta of the
decay products can also be used for the purpose [4,7].  In
principle, the effective dipole couplings can be complex.
Thus, measurements of CP asymmetries which are T-odd (like the
decay lepton energy asymmetry) give information on the real
parts of the dipole couplings, whereas those which are T-even
(like the decay-lepton up-down asymmetry about the production
plane) can measure the imaginary parts.  However, at
high-energies required for top production, when both $\g$ and $Z$
propagators are comparable, one still has to try and
disentangle the electric ($c^{\g}_d$) and weak dipole ($c^Z_d$)
couplings from each other. This problem has not received much
attention, and is one of the issues we discuss here.  We suggest
that this needs either studying additional asymmetry parameters,
or better still, using polarized beams\footnote{The advantage of
using polarized beams in the context of momentum correlations of
$t,\bar{t}$ decay products was pointed out in [7].}.

The purpose of this note is two-fold: firstly, we extend the
analysis of Chang {\it et al.} [5] to study new asymmetries in
$e^+e^-\ra t\bar{t}$ 
which are useful to get information on $\cdg$ and $\cdz$ separately. 
Secondly, we propose the use of longitudinal electron
polarization, which would be available at linear colliders, to
help in disentangling 
$\cdg$ from $\cdz$.  The
sensitivity in that case is considerably improved, as we shall
see.  

The process we consider is
\be
e^+(p_e)\;+\;e^-(p_{\bar{e}})\;\ra\;t(p_t)\;+\;\bar{t}(p_{\bar{t}}),
\ee
with subsequent decay of either $t$ or $\bar{t}$ via the leptonic channel,
\bea
t\,(p_t)\;&\ra&\;b\:+\:l^+\,(p_{l^+})\:+\:\nu_l, \nonumber \\
\bar{t}\,(p_{\bar{t}})\;&\ra&\;\bar{b}\:+\:l^-\,(p_{l^-})\:+\:\bar{\nu}_l.
\eea
We have in mind a future $e^+\,e^-$ collider, the Next Linear
Collider (NLC), and for definiteness we will assume the centre of
mass (cm) energy to be 500 GeV.

We assume the top quark to have couplings to $\g$ and $Z$ to be
given by the vertex factor  $ie\Gamma_\mu^j$, where
\be
\Gamma_\mu^j\;=\;c_v^j\,\g_\mu\;+\;c_a^j\,\g_\mu\,\g_5\;+
\;\f{c_d^j}{2\,m_t}\,i\g_5\,(p_t\,-\,p_{\bar{t}})_{\mu},\;\;j\;=\;\g,Z,
\ee
with
\bea
\cvg&=&\f{2}{3},\:\;\;\cag\;=\;0, \nonumber \\
\cdz&=&\f{\left(\f{1}{4}-\f{2}{3}\,x_w\right)}{\sqrt{x_w\,(1-x_w)}},
 \\
\caz&=&-\f{1}{4\sqrt{x_w\,(1-x_w)}}, \nonumber
\eea
and $x_w=sin^2\theta_w$, $\theta_w$ being the weak mixing
angle
\footnote{We largely adopt the notation of Chang {\it et al.} [5], to
facilitate comparison.}.
We have assumed in (3) that the only addition to the SM
couplings $c^{{\g},Z}_{v,a}$ are the CP-violating electric and weak dipole
form factors, $e\cdg/m_t$ and $e\cdz/m_t$, which are
assumed small.  Use has also been made of the Dirac equation in
rewriting the usual dipole coupling
$\sigma_{\mu\,\nu}\;(p_t\,-\,p_{\bar{t}})^{\nu}\;\g_5$ as
$i\;\g_5\;(p_t\,-\,p_{\bar{t}})_{\mu}$, dropping small corrections
to the vector and axial-vector couplings.  We assume that there
is no CP violation in $t,\,\bar{t}$ decay.

Using (3) to leading order in the dipole couplings, we have
calculated the following leptonic asymmetries for
arbitrary longitudinal electron (positron) beam polarizations
$P_e$ ($P_{\bar{e}}$)\footnote{We have restricted ourselves to
asymmetries for which analytic expressions could be obtained.}:
\vskip .2cm
\n 1. The energy asymmetry [5]
\be
 A_
E(x)\;=\;\f{1}{\sigma}\,\left[\f{d\sigma}{dx\,(l^+)}\:-\:\f{d\sigma}
{dx\,(l^-)}\right],
\ee
between distributions of $l^+$ and $l^-$ at the same value of
$x\:=\:x(l^+)\:=\:x(l^-)\:=\:4\, E(l^{\pm})\!/\!\sqrt{s}$.
\vskip .2cm
\n 2. The up-down asymmetry [5] $A_{ud}\:=
\:\int_{-1}^{+1}\!A_{ud}(\theta)\,d\cos\theta$, 
where
\be
A_{ud}(\theta)=\f{1}{2\,\sigma}\left[\f{d\,\sigma(l^+,{\rm
up})} {d\,\cos\theta}-\f{d\,\sigma(l^+,{\rm
down})}{d\,\cos\theta}+\f{d\,\sigma(l^-,{\rm up})}{d\,\cos\theta}-
\f{d\,\sigma(l^-,{\rm down})}{d\,\cos\theta}\right],
\ee
Here up/down refers to
$(p_{l^{\pm}})_y\;\raisebox{-1.0ex}{$\stackrel{\textstyle>}{<}$}\;0,\;
\:(p_{l^{\pm}})_y$ being the $y$
component of $\vec{p}_{l^{\pm}}$ with respect to a coordinate system
chosen in the $e^+\,e^-$ center-of-mass (cm) frame so that the
$z$-axis is along $\vec{p}_t$, and the $y$-axis is along
$\vec{p}_e\,\times\,\vec{p}_t$.  The $t\bar{t}$ production
plane is thus the $xz$ plane.  $\theta$ refers to the angle
between $\vec{p}_t$ and $\vec{p}_e$ in the cm frame.
\vskip .2cm
\n 3. The combined up-down and forward-backward asymmetry:
\be
A_{ud}^{fb}\;=\;\int_{0}^{1}\!A_{ud}(\theta)\,d\cos\theta\:-\:
\int_{-1}^{0}\!A_{ud}(\theta)\,d\cos\theta, 
\ee
with $A_{ud}(\theta)$ given in (6).
\vskip .2cm
\n 4. The left-right asymmetry
$A_{lr}\;=\;\int_{-1}^{+1}\!A_{lr}(\theta)\:d\cos \theta$, where
\be
A_{lr}(\theta)=\f{1}{2\,\sigma}\left[\f{d\,\sigma(l^+,{\rm
left})}{d\,\cos \theta}-\f{d\,\sigma(l^+,{\rm right})}
{d\,\cos \theta}+\f{d\,\sigma(l^-,{\rm left})}{d\,\cos
\theta}-\f{d\,\sigma(l^-,{\rm right})}{d\,\cos \theta}\right], 
\ee
Here left/right refers to
$(p_{l^{\pm}})_x\:\raisebox{-1.0ex}{$\stackrel{\textstyle>}{<}$} \:0$.
\vskip .2cm
\n 5. The combined left-right and forward-backward asymmetry
\be
A_{lr}^{fb}\;=\;\int_{0}^{1}\!A_{lr}(\theta)\,d\cos \theta\:-\:\int_{-1}^{0}\!A_{lr}(\theta)\,d\cos \theta,
\ee
with $A_{lr}(\theta)$ given in (8).

All these asymmetries are a measure of CP violation in the unpolarized
case and in the case when polarization is present, but
$P_e=-P_{\bar{e}}$.  When $P_e\neq -P_{\bar{e}}$, the initial
state is not invariant under CP, and therefore CP-invariant
interactions can contribute to the asymmetries.  However, to
the leading order in $\alpha$, these CP-violating contributions vanish
in the limit $m_e=0$.  Order-$\alpha$ collinear
helicity-flip photon emission can give a CP-invariant contribution to the
T-even asymmetries $(A_E, A_{lr}, A_{lr}^{fb})$.  However, this
background can be suppressed by a suitable cut on the visible
energy. The T-odd asymmetries $A_{ud}$ and $A_{ud}^{fb}$ are
genuine measures of CP violation even to order $\alpha$, since
an absorptive part, required for a CP-even and T-odd amplitude,
is not possible to that order.

Of these asymmetries, the energy asymmetry and up-down asymmetry
were discussed by Chang {\it et al.} [5].  The remaining are new, and
as we shall see, they help to measure different combinations of
${\rm Re}\,c_d^{\g}$ and ${\rm Re}\,c_d^Z$ or ${\rm Im}\,c_d^{\g}$ and ${\rm Im}\,c_d^Z$, thus
making possible a complete determination of all the parameters.
$A_E$ alone can only give one of ${\rm Im}\,c_d^{\g}$ or ${\rm Im}\,c_d^Z$,
assuming the other to be zero (or known).  A similar statement
is true of $A_{ud}$ and ${\rm Re}\,c_d^{{\g},Z}$. 

We give below expressions for the asymmetries in terms of
$c_d^i$\footnote{Our expression for the up-down asymmetry
agrees with that given in the errata of ref. [5].}:

\bea
A_E(x)&=&\f{2\,\beta}{C}\left\{f_L(x,\beta)-f_R(x,\beta)\right\}
\nonumber \\
& &\times\,\left\{{\rm Im}\,c_d^{\g}\:\left[(1-P_e\,P_{\bar{e}})\,
\left(2\,c_v^{\g}\,+\,(r_L+r_R)\,c_v^Z\right) 
\right. \right. \nonumber \\
 & &+\left.\,(P_{\bar{e}}-P_e)\,(r_L-r_R)\,c_v^Z\right] \nonumber \\
 & &+\,{\rm Im}\,c_d^Z\left[(1-P_e\,P_{\bar{e}})\,
\left((r_L+r_R)\,c_v^{\g}\,+\,(r_L^2+r_R^2)\,c_v^Z\right) 
\right. \nonumber \\
 & &+\left.
\,(P_{\bar{e}}-P_e)\,\left((r_L-r_R)\,c_v^{{\g}}\,+\,(r_L^2-r_R^2)\,
c_v^Z\right)\right\},
\eea
where
\bea
C&=&(1-P_e\,P_{\bar{e}})\,\left\{(3-\beta^2)\,
\left[\left(c_v^{{\g}}+r_Lc_v^Z\right)^2\,+\,
\left(c_v^{{\g}}+r_Rc_v^Z\right)^2\right] 
\right.  \nonumber \\
& &+\left. \,2\beta^2(c_a^Z)^2\,\left(r_L^2+r_R^2\right)\right\}
\nonumber \\
&
&+\,(P_{\bar{e}}-P_e)\,\left\{(3-\beta^2)\,\left[
\left(c_v^{{\g}}+r_Lc_v^Z\right)^2\,-\,\left(c_v^{{\g}}+
r_Rc_v^Z\right)^2\right] 
 \right. \nonumber \\
& &+\left. \,2\beta^2(c_a^Z)^2\,\left(r_L^2-r_R^2\right)\right\},
\eea
and the lepton energy distribution in $t$ decay is given for left
and right top helicities by [9]
\be
f_{L,R}(x,\beta)=\int_{\f{x}{1+\beta}}^{\f{x}{1-\beta}}\!f(x_0)\,
\f{\beta\,x_0\mp(x-x_0)}{2\,x_0^2\,\beta^2}\,dx_0,
\ee
$f(x_0)$ being the distribution in the $t$ rest frame,
\be
f(x_0)=\f{x_0(1-x_0)}{\f{1}{6}\,-\,\f{1}{2}\left(\f{m_W}{m_t}\right)^4\,+\,
\f{1}{3}\left(\f{m_W}{m_t}\right)^6}\;\,\theta(1-x_0)\;\theta
(x_0-\f{m_W^2}{m_t^2}).
\ee
$\beta$ is the top velocity in the cm frame,
$\beta=\left(1-4m_t^2/s\right)^{\f{1}{2}}$, and
$-er_{L,R}/s$ is the product of the $Z$-propagator and
left-handed (right-handed) electron couplings to $Z$, with
\bea
r_L&=&\f{\left(\f{1}{2}-x_w\right)}{\left(1-\f{m_Z^2}{s}\right)
\,\sqrt{x_w\,(1-x_w)}},
\nonumber \\
r_R&=&\f{-x_w}{\left(1-\f{m_Z^2}{s}\right)\,\sqrt{x_w\,(1-x_w)}}.
\eea

\bea
A_{ud}&=&-\f{3\pi\beta\sqrt{s}}{16m_tC}\left\{{\rm
Re}\,c_d^{{\g}}\left[(1-P_eP_{\bar{e}})(r_L-r_R)c_v^Z \right.
\right. \nonumber\\
&&\left. \left. +(P_{\bar{e}}-P_e)(2c_v^{{\g}}+(r_L+r_R)c_v^Z)\right]
\right. \nonumber \\
 & &\left. +{\rm Re}\,c_d^Z\left[(1-P_eP_{\bar{e}})
\left((r_L-r_R)c_v^{{\g}} +(r_L^2-r_R^2)c_v^Z\right)\right.
\right. \nonumber \\
& &+\left. \left. (P_{\bar{e}}-P_e)\left((r_L+r_R)
c_v^{{\g}}+(r_L^2+r_R^2)c_v^Z\right)\right]\right\} ,
\\[2mm]
A_{ud}^{fb}&=&\f{\beta^2\sqrt{s}}{4m_tC}c_a^Z\left\{{\rm
Re}\,c_d^{{\g}}\left[(1-P_eP_{\bar{e}})(r_L+r_R)\right. \right.
\nonumber \\
&& \left. \left. +(P_{\bar{e}}-P_e)(r_L-r_R)\right] 
\right. \nonumber \\
 &&+\left. {\rm
Re}\,c_d^Z\left[(1-P_eP_{\bar{e}})(r_L^2+r_R^2)+(P_{\bar{e}}-P_e)
(r_L^2-r_R^2)\right]\right\},\\[2mm]
A_{lr}&=&-\f{3\pi\beta^2\sqrt{s}}{16m_tC}c_a^Z\left\{{\rm
Im}\,c_d^{{\g}}\left[(1-P_eP_{\bar{e}})(r_L-r_R)\right. \right.
\nonumber\\ 
&& \left. \left. +(P_{\bar{e}}-P_e)(r_L+r_R)\right] 
\right. \nonumber \\
 &&+\left. {\rm
Im}\,c_d^Z\left[(1-P_eP_{\bar{e}})(r_L^2-r_R^2)+(P_{\bar{e}}-P_e)
(r_L^2+r_R^2)\right]\right\},\\[2mm]
A_{lr}^{fb}&=&\f{\beta\sqrt{s}}{4m_tC}\left\{{\rm
Im}\,c_d^{{\g}}\left[(1-P_eP_{\bar{e}})(2c_v^{{\g}}+
(r_L+r_R)c_v^Z)\right. \right. \nonumber \\ 
&& \left. \left. +(P_{\bar{e}}-P_e)(r_L-r_R)c_v^Z\right] 
\right. \nonumber \\
 &&+{\rm Re}\,c_d^Z\left[(1-P_eP_{\bar{e}})\left((r_L+r_R)
c_v^{{\g}}+(r_L^2+r_R^2)c_v^Z\right) 
\right. \nonumber \\
& &+\left. \left. (P_{\bar{e}}-P_e)\left((r_L-r_R)
c_v^{{\g}}+(r_L^2-r_R^2)c_v^Z\right)\right]\right\} .
\eea

Let us first look at the unpolarized case.  Whereas a
measurement of $A_E(x)$ would determine only one combination of
${\rm Im}\,c_d^{\g}$ and ${\rm Im}\,c_d^Z$, $A_{lr}$ and$A_{lr}^{fb}$ could be
used to determine two independent combinations, as can be seen
from (17) and (18) (putting $P_e\:=\:P_{\bar{e}}\:=\:0$).  These
could give ${\rm Im}\,c_d^{\g}$ and ${\rm Im}\,c_d^Z$ independently.
Similarly, measuring $A_{ud}^{fb}$ in addition to $A_{ud}$ would
help determine ${\rm Re}\,c_d^{\g}$ and ${\rm Re}\,c_d^Z$ independently.
Using a single asymmetry can only determine a combination of
dipole couplings, and in the absence of extra theoretical input
cannot say anything about individual dipole couplings.

However, if $e^-$ and $e^+$ beams have longitudinal
polarization, measuring the value of an asymmetry for two
different polarizations gives us two different combinations of
dipole couplings, allowing us to disentangle them.  It is
sufficient to measure for two different polarizations any one
CPT-even asymmetry to deduce both ${\rm Re}\,c_d^{\g}$ and ${\rm Re}\,c_d^Z$,
and any one CPT-odd asymmetry to deduce ${\rm Im}\,c_d^{\g}$ and ${\rm Im}\,c_d^Z$.

The two methods described above could have different
sensitivities.  Moreover, the sensitivity would be dependent on
which asymmetries are chosen within each method, and we will
later discuss our results in this context.  The measure of
sensitivity we use is the 90\% confidence level (C.L.) limit on
the dipole couplings 
which can be derived from a sample of appropriate events, and is
given by
\be
\delta c_d^i=\f{1.64\,\sqrt{N}\,c_d^i}{A},
\ee
where $A/c_d^i$ is the asymmetry for unit $c_d^i$ (assuming the
other $c_d^i=0$). When limits on both $c_d^{\gamma}$ and $c_d^Z$
are obtained independently, the factor 1.64 above is replaced by
2.15, corresponding to a 90\% C.L. limit for two degrees of freedom.

We also find that in some cases the sensitivity is greatly
enhanced by isolating the polarization-dependent part of the
distribution.  Thus, if we take a polarization asymmetrized
sample, corresponding to
$|d\sigma(P_e,P_{\bar{e}})-d\sigma(-P_e,-P_{\bar{e}})|$,
and evaluate all asymmetries with respect to this new sample, we
get a different set of asymmetries with different sensitivities.

We now come to our numerical results.  We assume that the NLC
will have 
$\sqrt{s}\:=\:500$ GeV and it is possible to have an integrated
luminosity of $\int\!{\cal L}\:=\:10\,{\rm fb}^{-1}$.
We take $m_t\:=\:174$ GeV and $x_w=.23$.  We look at only semi-leptonic
events, viz., when either of $t$ or $\bar{t}$ decays leptonically, while
the other decays hadronically.

Fig. 1 shows bands in the
${\rm Re}\,c_d^{\g}-{\rm Re}\,c_d^Z$ space which correspond to 2.15
$\sigma$ limit (90 \% C.L. for two degrees of freedom) obtained
from $A_{ud}$ and 
$A_{ud}^{fb}$, without polarization ($P_e\:=\:0$).  While $A_{ud}$ or
$A_{ud}^{fb}$ taken singly can limit one of ${\rm Re}\,c_d^{\g}$ or
${\rm Re}\,c_d^Z$ when the other is known, both the asymmetries put
together can provide independent limits on $|{\rm Re}\,c_d^{\g}|$ and
$|{\rm Re}\,c_d^{\g}|$, of the order of 5 and 1.5 respectively, for
$P_e\:=\:0$.  Also shown in Fig. 1 are the bands from $A_{ud}$ for
$e^-$ polarization $P_e\:=\:\pm\,0.5$ (with $P_{\bar{e}}\:=\:0$).
The limits obtainable are improved by an order of magnitude. 

Fig. 2 is in the ${\rm Im}\,c_d^{\g}-{\rm Im}\,c_d^Z$ space, and gives 2.15
$\sigma$ limits obtained from $A_{lr}$ and $A_{lr}^{fb}$.  Again,
for $P_e\:=\:0$, only a simultaneous search for both these
asymmetries can put independent limits on $|{\rm Im}\,c_d^{\g}|$ and
$|{\rm Im}\,c_d^Z|$, of the order of 0.7 and 6, respectively.
Limits on $A_{lr}$ with $P_e\:=\:\pm\,0.5$, also shown in Fig.
2, can improve these numbers by a
factor of about $5-7$. 

In Figs. 1 and 2, we do not show the combined up-down
(left-right) and forward-backward asymmetries for non-zero
polarization for clarity. However, the effect of polarization in
those cases is similar.

Thus, by using polarization, one can obtain independent limits of
the order of 0.2-0.25 on
three of the four dipole coupling parameters. The remaining
parameter, ${\rm Im}\,c_d^Z$ can be constrained to about 0.8.

Having considered independent limits, we now consider limits
obtained on either the electric or the weak dipole moment,
assuming the other dipole moment to be zero. 

For the energy asymmetry, we have estimated limits obtainable by
fitting the asymmetry in the range $x=0.1-1.5$. The improvement in
sensitivity is about a factor about 3 for $P_e=-0.5$ as
compared to $P_e=0$ for $|{\rm Im}\,c_d^Z|$, whereas measurement
of $|{\rm Im}\,c_d^{\g}|$ is insensitive to polarization.  However, on
considering a polarization asymmetrized sample (as described
above), the sensitivity for $|{\rm Im}\,c_d^Z|$ is improved by a
factor of about 12 as compared to the unpolarized case, giving
the best attainable 90\% C.L. limit as 0.06. 

The polarization asymmetrized distributions for $P_e=0.5$ leads
to an improvement in the sensitivities from the measurement of
$A_{ud}$ and $A_{ud}^{fb}$, whereas the sensitivity is worse in
the case of $A_{lr}$ and $A_{lr}^{fb}$. For example, $A_{ud}$
can give a limit on ${\rm Re}\,c_d^{\gamma}$ of 0.04 as compared
to 0.10 obtained without the asymmetrization procedure. 

One can also consider combinations of the different procedures
mentioned above to maximize the sensitivity available. We do not
enter into these details in this short note, since our purpose
is merely to point out the advantage that beam polarization can afford.

We end with the conclusions and a discussion of issues needing
further study.

We have calculated several CP-violating asymmetries which can arise in
the process $e^+\,e^-\;\ra\;t\,\bar{t}$, with subsequent
$t,\bar{t}$ decay, in the presence of electric and weak dipole
couplings of the top quark.  In order to disentangle the CP-violating
dipole couplings from each other, at least two T-odd asymmetries
are needed for the real parts and two T-even asymmetries are
needed for the imaginary parts, and we calculated possible
asymmetries which could be used for the purpose.  It
was shown that longitudinal 
polarization of the electron can help in separating the various
parameters, and in addition, leads to higher sensitivity.  At the NLC with
$\sqrt{s}\:=\:500\,GeV$ and polarized electron beams with
$\pm\,50\%$ polarization, 90\% C.L. sensitivities of the order
of 0.25 are obtainable on independent determinations  $|{\rm Re}\,c_d^{\g}|$,
$|{\rm Re}\,c_d^Z|$, and $|{\rm Im}\,c_d^{\g}|$, respectively, and a
sensitivitiy of about 0.8 for $|{\rm Im}\,c_d^Z|$.

Of these, the measurements of the real parts of $c_d^{{\g},Z}$ are
free from CP-invariant background contributions.  As for the T-even
asymmetries depending on the imaginary parts of $c_d^{{\g},Z}$,
the backgrounds from order-$\alpha$ collinear initial-state
photon emission have to be calculated and subtracted.  This will
be treated elsewhere.  As mentioned earlier, the background can
be reduced by imposing a cut on the visible energy. 

The theoretical predictions for $c_d^{\g,Z}$ are at the level of
$10^{-2}-10^{-3} $, as for example, in the Higgs-exchange
and supersymmetric models of CP violation.  Hence the measurements
suggested here cannot exclude these models at the 90\% C.L.
However, as simultaneous model-independent limits on both
$c_d^{Z}$ and $c_d^{\g}$, the ones obtainable from the
experiments we suggest, are an improvement over those obtainable
from measurements in unpolarized experiments.

Increase in polarization beyond $\pm 0.5$ can substantially
increase the asymmetries in some cases we consider.  Also, the
$e^+\,e^-$ cm energy can also have an effect on the asymmetries.
 A discussion of these effects is relegated to a future
publication.

We have compared our results with those of [7], where CP-odd 
momentum correlations are studied in the presence of $e^-$ 
polarization.  With comparable parameters, the sensitivities we
obtain are comparable to those obtained in [7].  In some cases
our sensitivities are slightly worse because we require either 
$t$ or $\bar{t}$ to decay leptonically, leading to a reduced event rate.
However, the better experimental efficiencies in lepton momentum
measurement may compensate for this loss.

Inclusion of experimental detection efficiencies may change our
results somewhat.  However, the main thrust of our conclusions,
that longitudinal beam polarization improves the sensitivity,
would still be valid.

One of us (S.D.R.) thanks Wai-Yee Keung for correspondence and clarifications
concerning ref. [5].

\newpage
\begin{center}
{\large \bf References}
\end{center}
\vskip .25cm
\noindent [1] CDF Collaboration, F.~Abe {\it et al.}, Phys.~Rev.~
Lett.~73 (1994) 225; Phys.~Rev.~D (to appear).

\noindent [2] I. Bigi and H. Krasemann, Z. Phys. C 7 (1981) 127;
J. K\"uhn, Acta Phys. Austr. Suppl. XXIV (1982) 203; I.
Bigi {\it et al.}, Phys. Lett. B 181 (1986) 157.

\noindent [3] 
J.F. Donoghue and G. Valencia, Phys. Rev.
Lett. 58 (1987) 451; C.A. Nelson, Phys. Rev. D 41 (1990) 2805;
G.L. Kane, G.A. Ladinsky and C.-P. Yuan, Phys. Rev. D 45 (1991) 124;
C.R. Schmidt and M.E. Peskin, Phys. Rev. Lett.
69 (1992) 410; C.R. Schmidt, Phys. Lett. B 293 (1992) 111; T. Arens and L.M.
Sehgal, Aachen preprint PITHA 94/14 (1994).

\noindent [4] W. Bernreuther, T. Schr\"oder and T.N. Pham,
Phys. Lett. B 279 (1992) 389; W. Bernreuther and P.
Overmann, Nucl. Phys. B 388 (1992) 53, Heidelberg
preprint HD-THEP-93-11 (1993); W. Bernreuther and
A. Brandenburg, Phys. Lett. B 314 (1993) 104; J.P.
Ma and A. Brandenburg, Z. Phys. C 56 (1992) 97; A.
Brandenburg and J.P. Ma, Phys. Lett. B 298 (1993) 211; W. 
Bernreuther and A.~Brandenburg, Phys.~Rev.~D {\bf 49}, (1994) 4481.

\noindent [5] D. Chang, W.-Y. Keung and I. Phillips, Nucl. Phys. B 
408 (1993) 286; 429 (1994) 255 (E).

\noindent [6] D. Atwood and A. Soni, Phys. Rev. D 45 (1992)
2405; B. Grzadkowski, Phys. Lett. B 305 (1993) 384; E. Christova
and M. Fabbrichesi, CERN preprint CERN-TH.6751/92, Revised
version (1993); A. Pilaftsis
and M. Nowakowski, Int. J. Mod. Phys. 9 (1994) 1097.

\noindent [7] F. Cuypers and S.D. Rindani, Munich preprint
MPI-PhT/94-54 (1994).

\noindent [8] B. Grzadkowski and W.-Y. Keung, Phys. Lett. B 316
(1993) 137; E. Christova and M. Fabbrichesi, CERN preprint
CERN-TH.6927/93 (1993).

\noindent [9] M. Je\'zabek and J.H. K\"uhn, Nucl. Phys. B320
(1989) 20.
\newpage
\begin{center}
{\large \bf Figure Captions}
\end{center}
\vskip .5cm
\n Fig. 1. Bands showing 90\% C.L. on the ${\rm Re}\,c_d^{\gamma}$
and ${\rm Re}\,c_d^Z$ coming from $A_{ud}$ (solid lines) and
$A_{ud}^{fb}$ (long-dashed lines) in the unpolarized case, and
from $A_{ud}$ with electron polarization $P_e=-0.5$
(short-dashed lines) and $P_e=+0.5$ (dotted lines).

\n Fig. 2. Bands showing 90\% C.L. on the ${\rm Im}\,c_d^{\gamma}$
and ${\rm Im}\,c_d^Z$ coming from $A_{lr}$ (solid lines) and
$A_{lr}^{fb}$ (long-dashed lines) in the unpolarized case, and
from $A_{lr}$ with electron polarization $P_e=-0.5$
(short-dashed lines) and $P_e=+0.5$ (dotted lines).

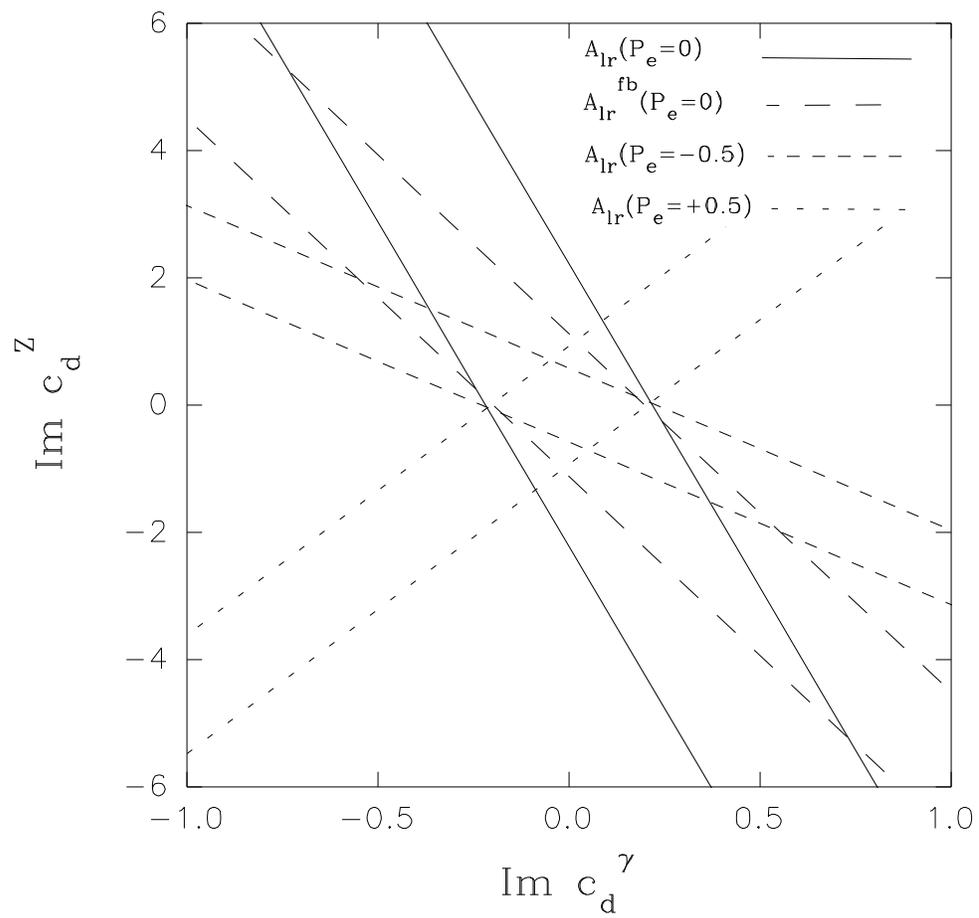
\end{document}